# Determining Which Sine Wave Frequencies Correspond to Signal and Which Correspond to Noise in Eye-Tracking Time-Series

Mehedi H. Raju*
Department of Computer Science
Texas State University
San Marcos, Texas, USA

Lee Friedman
Department of Computer Science
Texas State University
San Marcos, Texas, USA

Troy M. Bouman
Department of Mechanical
Engineering-Engineering Mechanics
Michigan Technological University
Houghton, MI, USA

Oleg V. Komogortsev
Department of Computer Science
Texas State University
San Marcos, Texas, USA

The Fourier theorem states that any time-series can be decomposed into a set of sinusoidal frequencies, each with its own phase and amplitude. The literature suggests that some frequencies are important to reproduce key qualities of eye-movements ("signal") and some of frequencies are not important ("noise"). To investigate what is signal and what is noise, we analyzed our dataset in three ways: (1) visual inspection of plots of saccade, microsaccade and smooth pursuit exemplars; (2) analysis of the percentage of variance accounted for (PVAF) in 1,033 unfiltered saccade trajectories by each frequency band; (3) analyzing the main sequence relationship between saccade peak velocity and amplitude, based on a power law fit. Visual inspection suggested that frequencies up to 75 Hz are required to represent microsaccades. Our PVAF analysis indicated that signals in the 0-25 Hz band account for nearly 100% of the variance in saccade trajectories. Power law coefficients ($a, b$) return to unfiltered levels for signals low-pass filtered at 75 Hz or higher. We conclude that to maintain eye-movement signal and reduce noise, a cutoff frequency of 75 Hz is appropriate. We explain why, given this finding, a minimum sampling rate of 750 Hz is suggested.

## Introduction

Fourier analysis models a time-series as the sum of a set of sinewaves with variable frequencies, phases, and amplitudes. In many cases (but not all, e.g., nystagmus; Rosengren et al., 2020), the lower frequencies are required to preserve a time-series feature of interest (e.g., saccade peak velocity), and higher frequencies may not be needed and thus represent noise. A low-pass filter can effectively retain the signal part of the time-series while eliminating the noise part, making it a suitable solution for this typical scenario.

The best practice would be for researchers to evaluate what frequencies are needed, and which are not, for specific goal, prior to data collection. This preliminary analysis would allow the researcher to design an appropriate data collection scheme. In this part of the study, signals should be collected at the highest possible frequency so that an analysis of which frequencies are needed can be performed. Once this information is known, filter settings and sampling rate can be optimized.

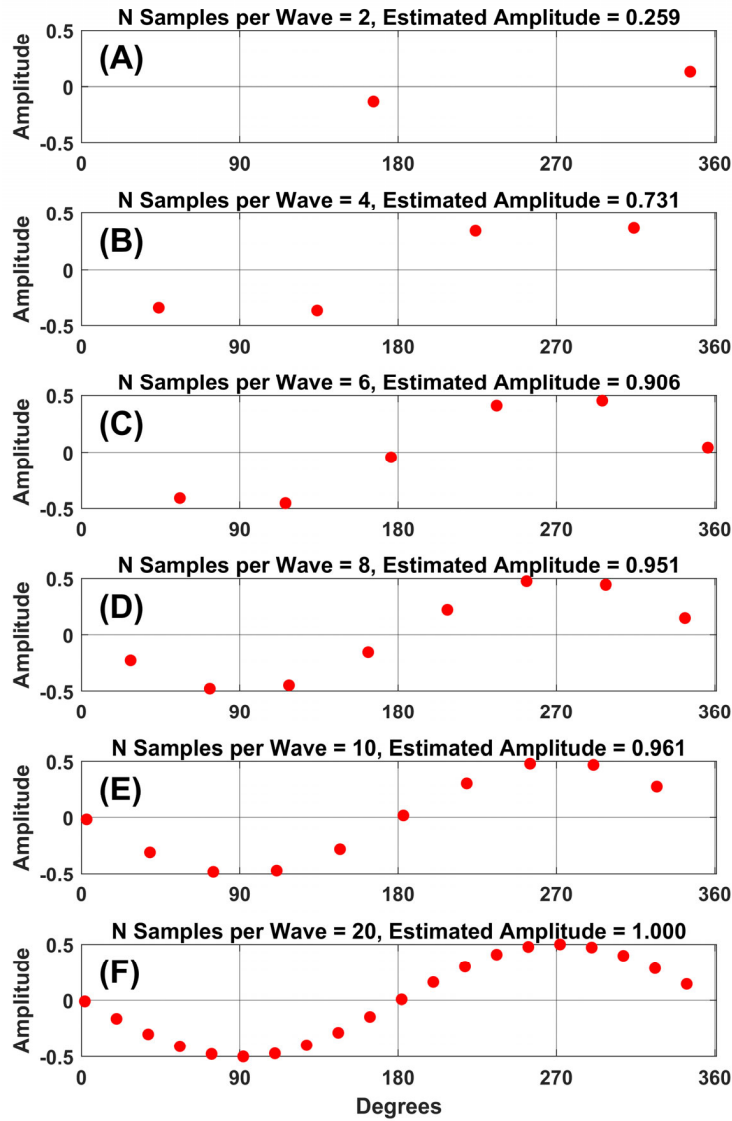

Figure 1. Visual representation of the 10x rule

The study goals are very important when trying to determine a required sampling rate. If we are interested in the frequency domain, then a minimum of 2 samples per wave is required (Shannon, 1949). If the fastest frequency we need was F Hz, then the minimum sampling frequency needs to be 2 x F Hz. However, if we are interested in the time domain, as we believe that most eye-movement researchers are, then the minimum sampling frequency (Fs) needs to be Fs >= 10 x F Hz. The signal processing basis for the 10x rule is discussed at the links associated with these references (Instruments; Siemens; WikiBooks). We are not aware of any published reference for this 10x rule. The need for 10x sampling is illustrated in Figure 1 and Figure 2. As is evident in Figure 1, a sine wave becomes less resolvable as the number of sampling rates per period decreases. If one's goal is to accurately measure the amplitude of a sinewave of a particular frequency, then Figure 2 illustrates that the estimates of amplitude can be very inaccurate with fewer than 10 samples per sine wave.

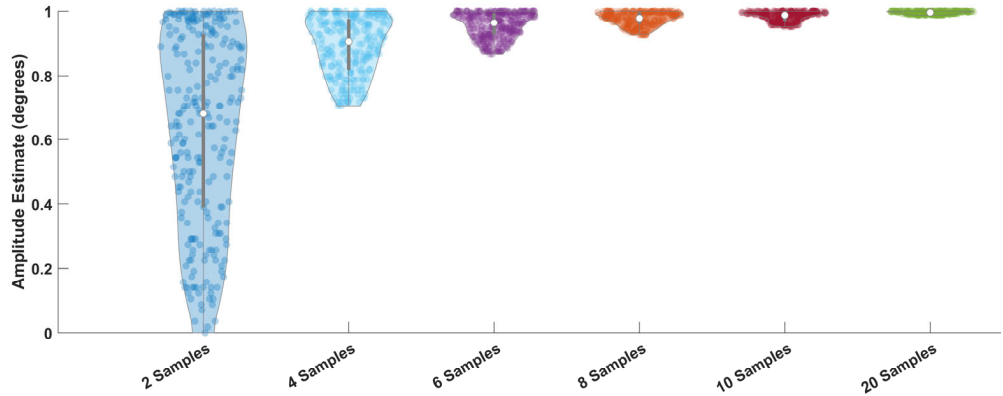

Figure 2: Violin plots of amplitude estimates of a sine wave with an amplitude of 1 deg.

We are not aware of any paper in the eye-movement field that used this 10x rule, including all the papers cited in this study. Here, a leading research group makes this statement:

> "For oscillating eye-movements, such as tremors, we can argue based on the Nyquist-Shannon sampling theorem (Shannon, 1949) that the sampling frequency should be at least twice the speed of the particular eye movement (e.g., behavior at 150 Hz requires >300 Hz sampling frequency)" (Andersson et al., 2010).

This statement is only true if the goal is a frequency domain analysis. Typically, eye movement researchers are interested in the pattern of eye-movement waveforms, e.g., the trajectories of saccades, or PSOs, or the length and stability of fixation, etc... Therefore, in most cases, the correct rule of thumb is the 10x rule described above.

Below, we review the prior research on required frequencies for saccades. For our research (and we suspected many others) faithful preservation of saccade trajectories and main-sequence-related saccade metrics would probably be sufficient. We didn't review signal-to-noise determinations for ocular microtremor, a high frequency component of fixation which cannot be measured with video-oculography (McCamy et al., 2013).

We present our analysis of the literature in Table 1. Two potentially relevant papers (Inchingolo & Spanio, 1985; Juhola et al., 1985) were not included in our table. In (Juhola et al., 1985), the signals (electrooculography EOG and photoelectric) were analog signals. These analog signals were filtered first with the low-pass analog filter at 30 Hz. Subsequently, the signals were digitally filtered with a low-pass filter with a cutoff of 70 Hz. This creates a very complex situation, and we didn't think that statements about frequencies required to preserve saccade peak velocity were useful given the insertion of this analog filter. In Inchingolo and Spanio (1985), the research was based on an EOG signal which was analog filtered with a cutoff at 100 Hz. Any further statements about the effects of other digital low-pass filtering were confounded by the presence of the analog filter. Therefore, these papers were not included in Table 1.

From Table 1, despite the difference in recording and other methods, the literature supports the notion that 0-125 Hz frequency components are sufficient to preserve saccade characteristics. Although there is literature relevant to this topic, we wanted to make our own determination using novel criteria not previously reported. Our goal in this study was to determine which frequencies are needed to preserve signal and which frequencies correspond to noise in eye-tracking studies. We evaluate this issue for saccades, microsaccades and smooth pursuit.

Table 1. Frequency Content of Saccades

| Citation | Methods | Findings |
| --- | --- | --- |
| **Bahill (1981)** | Photoelectric techniques | For noisy data, a bandwidth of 0-125 Hz was required to record saccades. Also, a sampling rate of 1000 Hz was suggested. |
| **Schmitt et al. (2007)** | Video-based infrared eye-tracker | Sampling rate should be 250 Hz. |
| **Wierts et al. (2008)** | VOG and Search coil | Saccadic eye movements of >=5° amplitude were bandwidth limited up to a frequency of 25 to 30 Hz. A sampling frequency of about 50 Hz was sufficiently high to prevent aliasing. |
| **Mack et al. (2017)** | Synthetic saccades | Signals sampled as low as 240 Hz allow for the good reconstruction of peak velocity. With 240 Hz, the frequencies that can be evaluated were 0-24 Hz (in the time domain). |

# Methods

## Subjects

We recorded a total of 23 unique subjects (M=17/ F=6, median age = 28, range = 20 to 69 years). Of the total number of unique participants, 14 had normal (not-corrected) vision, and 9 had corrected vision (7 glasses, 2 contact lenses). Nine of the unique participants were left-eye dominant and 14 were right-eye dominant. Eye dominance was determined using the Miles method (Miles, 1930). Subjects were recruited from laboratory personnel, undergraduates taking a class on computer programming, and friends of the experimenters. The Texas State University institutional review board approved the study, and participants provided informed consent.

We used two datasets. The first dataset, called "Fixation", originally had data from 15 subjects. However, due to blinks and other artifacts, we only analyzed data from 9 subjects. The second dataset, called "RS-SP", included data from 9 subjects who performed both a random saccade task and a smooth pursuit task.

## Eye Movement Data Collection

Eye movements were collected with a tower mounted EyeLink 1000 eye tracker (SR Research, Ottawa, Ontario, Canada). The eye tracker operated in monocular mode capturing the participant's dominant eye. During the collection of eye movements data, each participant's head was positioned at 550 millimeters from a 19" (48.26 cm) computer screen (474 x 297 millimeters, resolution 1680 x 1050 pixels), where the visual stimulus was presented. The sampling rate was 1000 Hz. All datasets were collected with all heuristic filters off, i.e., unfiltered.

For the fixation task, subjects were presented with a single fixation point (white circle, 0.93°) as the visual stimulus. The point was positioned in the horizontal middle of the screen and at a vertical angle of 3.5° above the primary position (position of the eye when looking straight ahead). Participants were instructed to fixate on the stationary point stimulus for a period of 30 seconds (Griffith et al., 2021; Raju, 2022).

During the random saccade task, subjects were instructed to follow the same target on a dark screen as the target was displaced at random locations across the display monitor, ranging from ± 15º and ± 9º of visual angle in the horizontal and vertical directions respectively. The random saccade task was 30 seconds long. The target positions were randomized for each recording. The minimum amplitude between adjacent target displacements was 2º of visual angle. The distribution of target locations was chosen to ensure uniform coverage across the display. The delay between target jumps varied between 1 sec and 1.5 sec (chosen randomly from a uniform distribution).

During the smooth pursuit task, subjects were instructed to follow a target on the dark screen as the target moved horizontally from center to right. This ramp was followed by a fixation (length between 1 and 1.5 sec). This was followed by another ramp from the right to the left of the screen, then another fixation, etc. The rest of the task was a series of left-to-right and right-to-left ramps with fixations interposed. The target was moving at velocities of either 5º/sec, 10º/sec, or 20º/sec. For each speed, there were 5 continuous leftward and 5 rightward ramps per set. The order of the velocity sets was random for each participant. There was a 15 sec fixation period at the beginning of the task and between each set. The whole recording was 120 seconds long.

### Selection of Saccade, Catch-up Saccade and Microsaccade Exemplars

We wanted to have multiple exemplars of saccades, catch-up saccades (CUS), and microsaccades. Catch-up saccades occur when tracking a smoothly moving target. When the gain was less than 1.0, subjects consistently lag the smoothly moving signal. In this case, they generate relatively small saccades to "catch-up" to the target. For a detailed analysis of the relationship between smooth pursuit gain, CUS amplitude, and CUS rate see (Friedman et al., 1991)

For the saccade examples, we used the random saccade task. For the CUS, we used the smooth pursuit dataset. Microsaccades were selected from the fixation dataset. Two exemplars were chosen for each eye movement type, a low-noise example, and a high-noise example ("clean" and "noisy"). The selection was subjective but incorporated measures of precision to guide this choice. More examples are available as part of our supplementary material.

### Signal Frequency Content Analysis

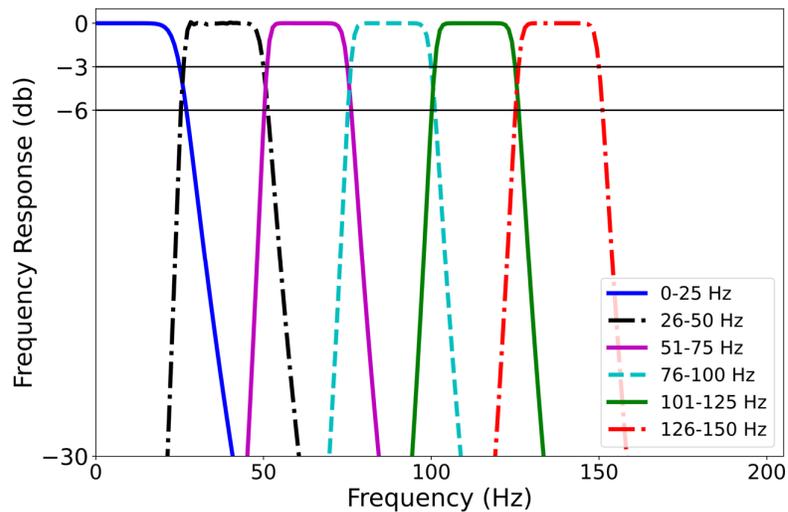

Figure 3. Frequency response of different frequency bandwidths using 7$^{th}$ order, zero-phase Butterworth filters

We wanted to evaluate eye movements after one of seven filtering regimes (unfiltered, low-pass filtered at 25 Hz, band pass filtered from 26-50 Hz, 51-75 Hz, 76-100 Hz, 101-125 Hz, and 126-150 Hz). See Figure 3 for an illustration of the frequency-response of the various filters. These were created using very sharp high-pass, low-pass, and band-pass Butterworth-style filters (order = 7). To prevent phase effects, all these filters were zero-phase, which means that after the data were filtered in the forward direction, the signal was flipped, and the signal passed through the filter again. This procedure effectively doubled the filters' orders and squares the magnitudes of their transfer functions. The filtering operation was performed in post-processing.

### Calculation of Percentage of Variance Accounted For (PVAF)

The first step for this analysis was to identify saccades in all our Random Saccade task data. The identification was initially performed by an updated version of our previously published event detection method (Friedman et al., 2018). All potential saccades were screened by the authors so that only well-marked saccades were included. There were 1,033 well-marked saccades (out of total 1910 saccades). A PVAF analysis was performed on each of these saccades.

For each of these saccades, data from the unfiltered condition was treated as a dependent variable, and all the filtered signals were treated as independent variables. We regressed the first filtered signal (0-25 Hz) onto the unfiltered signal and noted the $r^2$. We then added the data filtered from 26-50 Hz and noted the change in $r^2$. We kept doing this until all the filtered bands had been entered into the multiple linear regression model. In this case, we evaluated PVAF at 0-25, 26-50, 51-75, 76-100, 101-125 and 126-150 Hz bands. We multiplied each $r^2$ by 100 to obtain the percent of variance accounted for (PVAF).

### Study of the Effects of Filtering on the Main Sequence Relationship between Saccade Peak Velocity and Saccade Amplitude

The goal of this analysis was to evaluate the effects of low-pass filtering on the saccade main sequence relationship between horizontal saccade amplitude and horizontal peak velocity. In this analysis, the relationship was represented in each condition (unfiltered and filtered at 25, 50, 75, 100, 125 and 150 Hz) as a power law ($y=a*x^b$), where x is saccade amplitude, and y is peak velocity. Confidence limits (95%) were estimated for each coefficient. The question we ask is: How do the coefficients and their confidence limits in various filter conditions compare to the coefficient estimate in the unfiltered condition.

In addition to the power law relationships, we also tested an exponential fit suggested by (Leigh & Zee, 2015) (page 172):

$$\text{Peak Velocity} = V_{max} * (1 - \exp^{(-\text{Amplitude}/C)}) \qquad (1)$$

where Vmax is the asymptotic peak velocity and C is a constant.

However, in 6 of 7 cases, the adjusted model $r^2$ was higher for the power law fits than the exponential fits (Paired t-test, $t = 4.09$, $df = 6$, $p = 0.006$, two-tailed). Therefore, we only present results for the power law fits.

We started with the 1,033 saccades discussed above. Snippets of the horizontal position channel were cut from 200 ms prior to each saccade to 200 ms after each saccade. For each snippet, a velocity calculation was performed using the 1st derivative from a Savitzky-Golay filter function with order=2 and window=7 (Friedman et al., 2017). The peak (absolute) horizontal velocity and the absolute horizontal amplitude of each saccade was determined.

## Results

### Analysis of Exemplars

#### Saccades

In Figure 4, we present the signal frequency content analysis for a "clean" saccade. This saccade has an approximate amplitude of 2.94°. In plot (A1) we present the unfiltered signal trace for the saccade. In plots (B1) to (G1) we present the signal containing frequencies from different bands. All the plots in the left column were scaled to match the unfiltered saccade in (A1). All the plots on the right column were scaled individually based on their range of data. The signal in plot (B1) appears very similar to the saccade in plot (A1). However, the post-saccadic activity in (A1) was missing, and there was less noise. The saccade amplitude has not been altered.

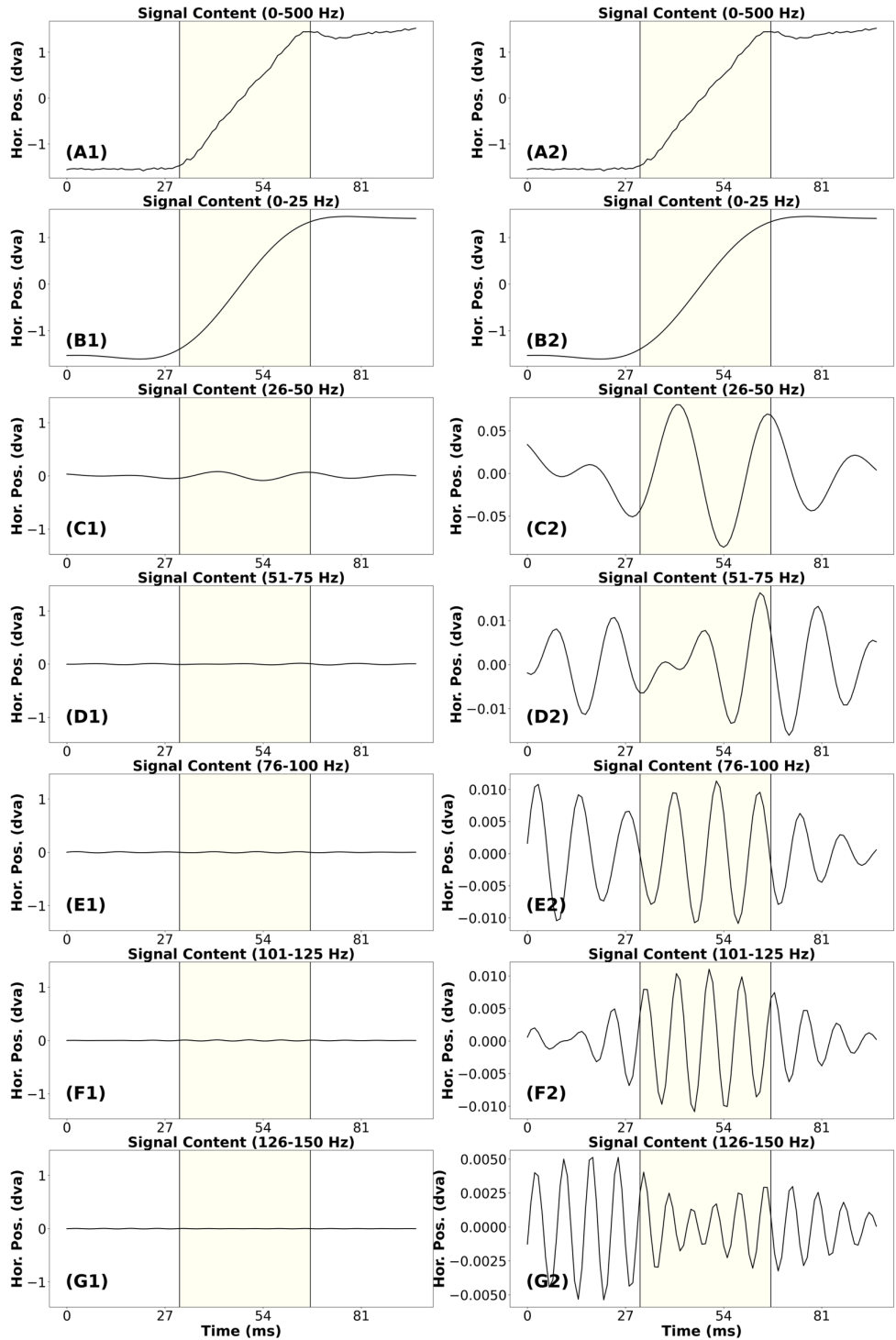

Figure 4. Signal frequency content analysis of a clean saccade. (A1) Exemplar of a clean unfiltered saccade. (B1) The signal in (A1) with only frequencies from 0 to 25 Hz. (C1) The signal in (A1) with only frequencies from 26 to 50 Hz. (D1) The signal in (A1) with only frequencies from 51 to 75 Hz. (E1) The signal in (A1) with only frequencies from 76 to 100 Hz. (F1) The signal in (A1) with only frequencies from 101 to 125 Hz.

(G1) The signal in (A1) with only frequencies from 126 to 150 Hz. Note that all plots on the left panel have the same amplitude range as the original saccade. Since we cannot see some of the signals on this scale very well, each plot (A2-G2) on the right panel was y-scaled individually according to the range of the data. Yellow highlighting indicates the saccade.

In plot (C1) we present the signal containing frequencies from 26-50 Hz. There appears to be a minor contribution to signal amplitude from this band. For the remaining plots in the left column (D1 to G1), it appears that no signal remains that was relevant to the trajectory of the unfiltered saccade in (A1). In the right column, note the range of the data in (D2) to (G2). All these bands contribute less than 3.0% of the amplitude of the unfiltered saccade. The waveforms of these plots do not appear to be relevant to the unfiltered saccade. So, for this saccade, we would consider that the data below 50 Hz were signal and the data above 50 Hz were noise.

In Appendix Figure 1, we present the signal frequency content analysis for a "noisy" saccade. This saccade has an approximate amplitude of 2.79°. In plot (A1) we present the unfiltered signal trace for the saccade. In plots (B1 to G1) we present the signal containing frequencies from different bands. The signal in plot (B1) appears very similar to the saccade in plot (A1). However, the post-saccadic activity in (A1) was missing, and there was less noise. The saccade amplitude has not been altered. In plot (C1) we present the signal containing frequencies from 26-50 Hz. There appears to be a minor contribution to signal amplitude from this band. Some of the signals in this band may contribute to the post-saccadic activity in the unfiltered saccade. For the remaining plots in the left column (D1 to G1), it appears that no signal remains that was relevant to the trajectory of the unfiltered saccade in (A1).

In the right column, note the range of the data in (D2) to (G2). All these bands contribute less than 4.0% of the amplitude of the unfiltered saccade. The waveforms of these plots do not appear to be relevant to the unfiltered saccade. So, for this saccade also, we would consider that the data below 50 Hz were signal and the data above 50 Hz were noise.

## Microsaccade

In Appendix Figure 2, we present the signal frequency content analysis for a "clean" microsaccade. This microsaccade has an approximate amplitude of 0.63°. The saccade detection algorithm determined the end of this saccade later than one would choose manually, but we don't think this difference affects the present analysis. In plot (A1) we present the unfiltered signal trace for the microsaccade. In plots (B1) to (G1) we present the signal containing frequencies from different bands. The signal in plot (B1) appears to be a very smooth version of the waveform in (A1). The microsaccade amplitude may be very slightly less than the amplitude of the unfiltered microsaccade. In plot (C1) we present the signal containing frequencies from 26-50 Hz. The waveform for the data filtered at 51-75 Hz (D1) appears to contain some relevant signal. For the remaining plots in the left column (E1 to G1), it appears that no signal remains that was relevant to the trajectory of the unfiltered microsaccade in (A1).

Similarly, in Appendix Figure 3, we present the signal frequency content analysis for a "noisy" microsaccade. This microsaccade has an approximate amplitude of 0.651°. The signal in plot (B1) looks like a very smooth version of the unfiltered saccade.

The amplitude of this very smooth waveform is, at most, very slightly less than the unfiltered saccade. In plot (C1), we present the signal containing frequencies from 26-50 Hz. These higher frequencies contribute to the sharpness of unfiltered signal. In plot (D1), we present the signal containing frequencies from 51-75 Hz. As was the case for the waveform in (C1), these higher frequencies contribute to the sharpness of the unfiltered signal. For the remaining plots in the left column (E1 to G1), it appears that no signal remains that was relevant to the trajectory of the unfiltered microsaccade in (A1). For the remaining plots in the left column (E1 to G1), it appears that no signal remains that was relevant to the trajectory of the unfiltered microsaccade in (A1). This part of the signal was what makes this a "noisy" saccade. For this microsaccade also, we would consider that the data below 75 Hz were signal and the data above 75 Hz were noise.

### Smooth Pursuit and Catch-up saccades (CUS)

In Appendix Figures 4 and 5, we present a "clean" and "noise" segment of smooth pursuit. Both segments have five or more CUS. The analysis of these figures was identical. In the (A1) plot we present the unfiltered smooth pursuit signal, including catch-up saccades. The (B1) plots appear very similar to that of the unfiltered segment. In the band from 26-50 Hz, there are some very small high-frequency bursts coincident with each saccade. Plot (C2) makes this point more clearly. For the remaining plots (D1 to G1), it appears that no signal remains that was relevant to the pattern of smooth pursuit of the unfiltered catch-up saccade in (A1). In both plots (D2) and (E2), there were bursts of high-frequency noise signal coincident with each CUS. However, the amplitude of these bursts in (E2) was so small that we think data from this frequency band can be ignored. For these smooth pursuit segments, we would consider that the data below 50 was signal and the data above 50 Hz was noise.

### Percentage of Variance Accounted For (PVAF)

Our results for the PVAF analysis of all saccade trajectories are presented in Figure 5. It was clear from this figure that nearly all the variance in the trajectory of the unfiltered saccade was accounted for by data in the range of 0-25 Hz. None of the high-frequency data contributed to the variance in the original unfiltered saccade in any substantial way. See Figure 5 for exact numbers.

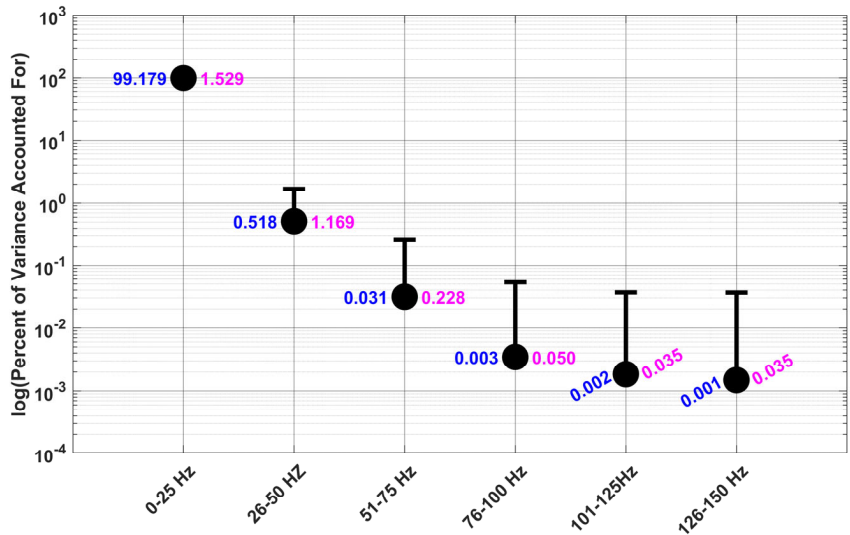

Figure 5. PVAF Analysis at different filtered levels. Each circle was the median PVAF (exact numbers in blue) for a particular filter level. Median absolute deviation (MAD) values for each point are in pink.

### Effects of Filtering on the Main Sequence Relationship between Saccade Peak Velocity and Saccade Amplitude

This main sequence relationship in the unfiltered condition is illustrated in Figure 6. In the next step, each snippet was filtered with 7$^{th}$ order (zero-phase) low-pass (zero-phase) Butterworth filter with cutoffs of 25, 50, 75, 100, 125 and 150 Hz. The main sequence relationships for these 6 conditions are illustrated in Figure 7. The model adjusted r-squares for all 7 models (unfiltered, 25 Hz, 50 Hz, 75 Hz, 100 Hz, 125 Hz and 150 Hz) ranged from 0.89 to 0.94.

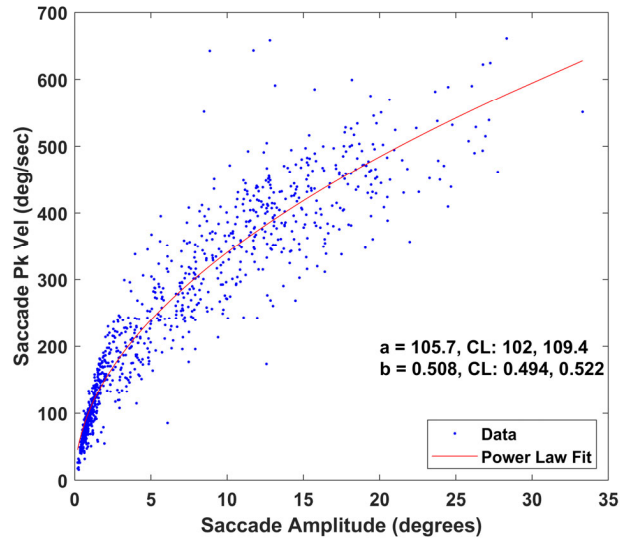

Figure 6. Main sequence relationship between horizontal saccade peak velocity and amplitude (N=1,033 saccades). Note the power law fit (y=a*x$^b$, x is horizontal amplitude, y is horizontal peak velocity, red line) and the *a* and *b* estimates. Also included are the 95% confidence limits (CL) for the estimates.

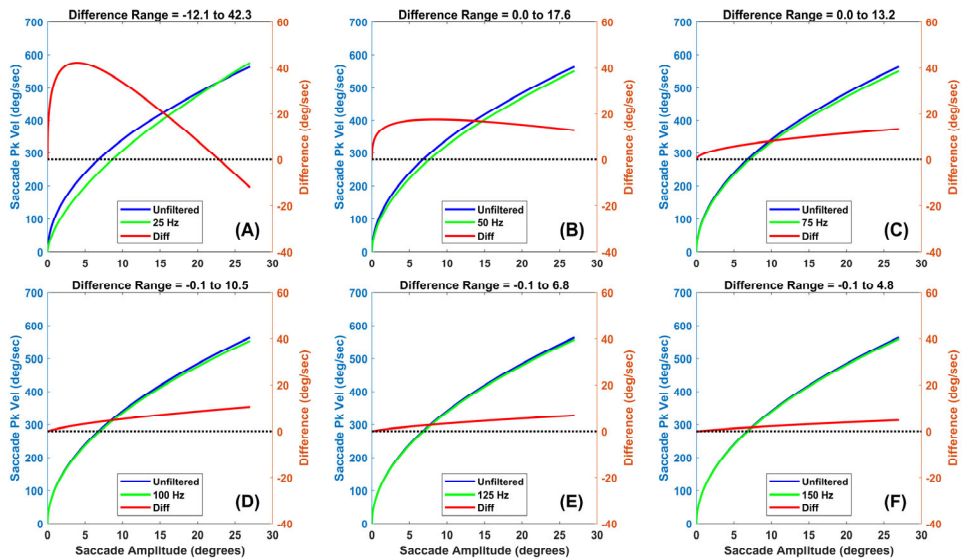

Figure 7. Main sequence power law fits (y=a*x$^b$, where x is horizontal saccade amplitude and y is horizontal peak velocity) for unfiltered data and data filtered at various levels. (A) The unfiltered power law relationship for unfiltered data (blue) and data filtered at 25 Hz (green) is illustrated. The left ordinate axis is in units of peak velocity. In red, we show the difference (unfiltered – filtered) with units on the right ordinate axis. (B) Same as (A) but data filtered at 50 Hz. (C) Same as (A) but data filtered at 75 Hz. (D) Same as (A) but data filtered at 100 Hz. (E) Same as (A) but data filtered at 125 Hz. (F) Same as (A) but data filtered at 150 Hz.

In Figure 7(A), the saccade peak velocity of the filtered signal is lower than in the unfiltered condition up to an amplitude near 26 degrees. Above this level, the estimated peak velocity of the filtered data is slightly higher than that of the unfiltered condition. For all other subplots in Figure 7, estimates of peak velocity in the filtered condition were always lower than the estimates when the data were unfiltered. So, low pass filtering tends to lower the estimate of saccade peak velocity. Considering a maximum peak velocity in the study near 600 deg/sec, the underestimates of saccade peak velocity are quite small for figures 7(C through F).

In Figure 8, we present the power law coefficients (*a* and *b*) and their 95% confidence limits in the unfiltered condition and at the various filter levels. For the *a* coefficient (Figure 8(A)), we can see that the 95% confidence limits at 75 Hz include the unfiltered *a* estimate. This means that the *a* coefficient estimates at 75 Hz, as well as the coefficients for 100, 125 and 150 Hz, are not statistically significantly different from that in the unfiltered condition.

Notice in Figure 8(B) that there is no significant difference between the *b* coefficient in the unfiltered condition and the *b* coefficient for any filter level from 75 to 150 Hz.

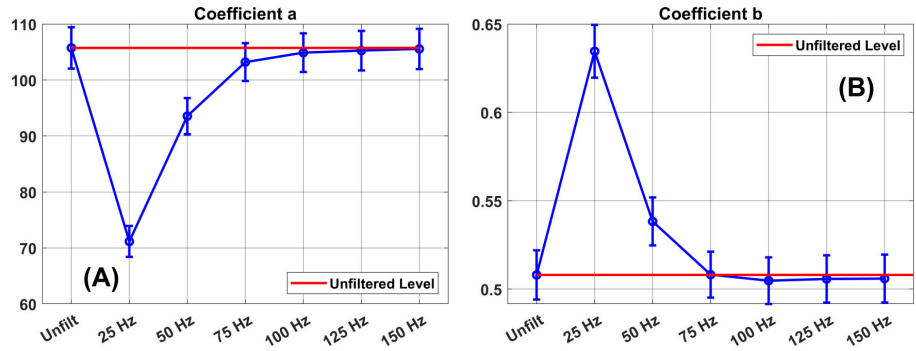

Figure 8: Power law coefficients and their 95% confidence limits. The *a* coefficient estimates in the unfiltered condition and in the several filter conditions are illustrated in (A). The error bars indicate the 95% confidence limit around each coefficient estimate. The *b* coefficient estimates in the unfiltered condition and in the several filter conditions are illustrated in (B).

## Summary of Results

In Table 2, we summarize our conclusions about which frequencies correspond to signal and which correspond to noise.

Table 2. Summary of Results

| Evidence | Method | What is Signal | What is Noise |
|---|---|---|---|
| **Exemplars** | Visual Inspection of Saccade | 0-50Hz | 51-500 Hz |
|  | Visual Inspection of Microsaccade | 0-75Hz | 76-500 Hz |
|  | Visual Inspection of CUS | 0-50Hz | 51-500 Hz |
| **Variance Explained** | Compute percent of variance accounted for in unfiltered saccades | 0-25 Hz | 26-500 Hz |
| **Main Sequence analysis** | Power Law Coefficients | 0-75 Hz | 76-500 Hz |

## Discussion

We have assessed the appropriate frequency cutoff between signal and noise using three approaches. The different analyses provide different answers but can be summarized in a final single rule. The visual analysis of our microsaccade and smooth pursuit exemplars suggested that frequencies up to 75 Hz were required to retain signal whereas frequency content above 75 Hz represent noise. Our analysis of the percent of variance accounted for in unfiltered saccade trajectories by different filter bands indicated that essentially all the variances in saccade shape were accounted for with data in the 0-25 Hz band. The power law coefficients (*a* and *b*) for the main sequence relationship between horizontal saccade peak velocity and horizontal saccade amplitude show no significant differences between the unfiltered condition and filtering conditions from 75 to 150 Hz. Taken together, we conclude that, if the goal is to preserve saccade (including microsaccade) and smooth pursuit characteristics, frequencies up to 75 Hz are required and frequencies above this are noise.

These results have implications for proposed sampling frequencies for future data collection. If our studies only involved the frequency domain processing, we would only need two samples at 75 Hz, so a sampling rate of 150 Hz would suffice. However, because we were interested in evaluating eye movements with time domain processing, the 10x rule discussed in the introduction applies. Therefore, the minimum acceptable sampling rate for eye-tracking studies is 750 Hz.

Our observations apply to data collected from an EyeLink 1000 eye tracker with a 1000 Hz sampling rate. Our observations apply to studies involving fixation, saccades, catch-up saccades, microsaccades and smooth pursuits. These results may be device- and eye movement type-specific. However, we do provide a general framework for addressing the question "What is signal and what is noise?" in any time-series dataset.

## Conclusion

In this paper, we tried to answer the vital question about recorded eye movements: Which sine-wave frequencies correspond to signal, and which correspond to noise? We employed several approaches to this problem. We conclude that frequencies up to 75 Hz are signal, and frequencies above that are noise. We explain why, if the interest is in a time-domain analysis, which we believe is the most relevant domain for eye-movement research, there is a 10x rule of thumb to determine the required sampling rate. With this rule of thumb, the required sampling rate to accurately represent sine-waves of 75 Hz is 750 Hz. In our follow-up to this article (Raju, 2023), we compare various filter schemes in terms of their ability to preserve signal and remove noise. Ultimately, we make recommendations for EyeLink users going forward.

## Data Availability Statement

As stated in the manuscript, the signals for all saccades analyzed are at https://digital.library.txstate.edu/handle/10877/16437. Also, images of all saccades as well as the basis for the PVAF analysis for each saccade are also available on this website.

## Ethics and Conflict of Interest

The author(s) declare(s) that the contents of the article are in agreement with the ethics described in http://biblio.unibe.ch/portale/elibrary/BOP/jemr/ethics.html and that there is no conflict of interest regarding the publication of this paper.

## Acknowledgements

This work was funded by a grant from the NSF (1714623) (PI: Oleg Komogortsev). The funders had no role in the design of the study; in the collection, analyses, or interpretation of data; in the writing of the manuscript, or in the decision to publish the results.

# Appendix

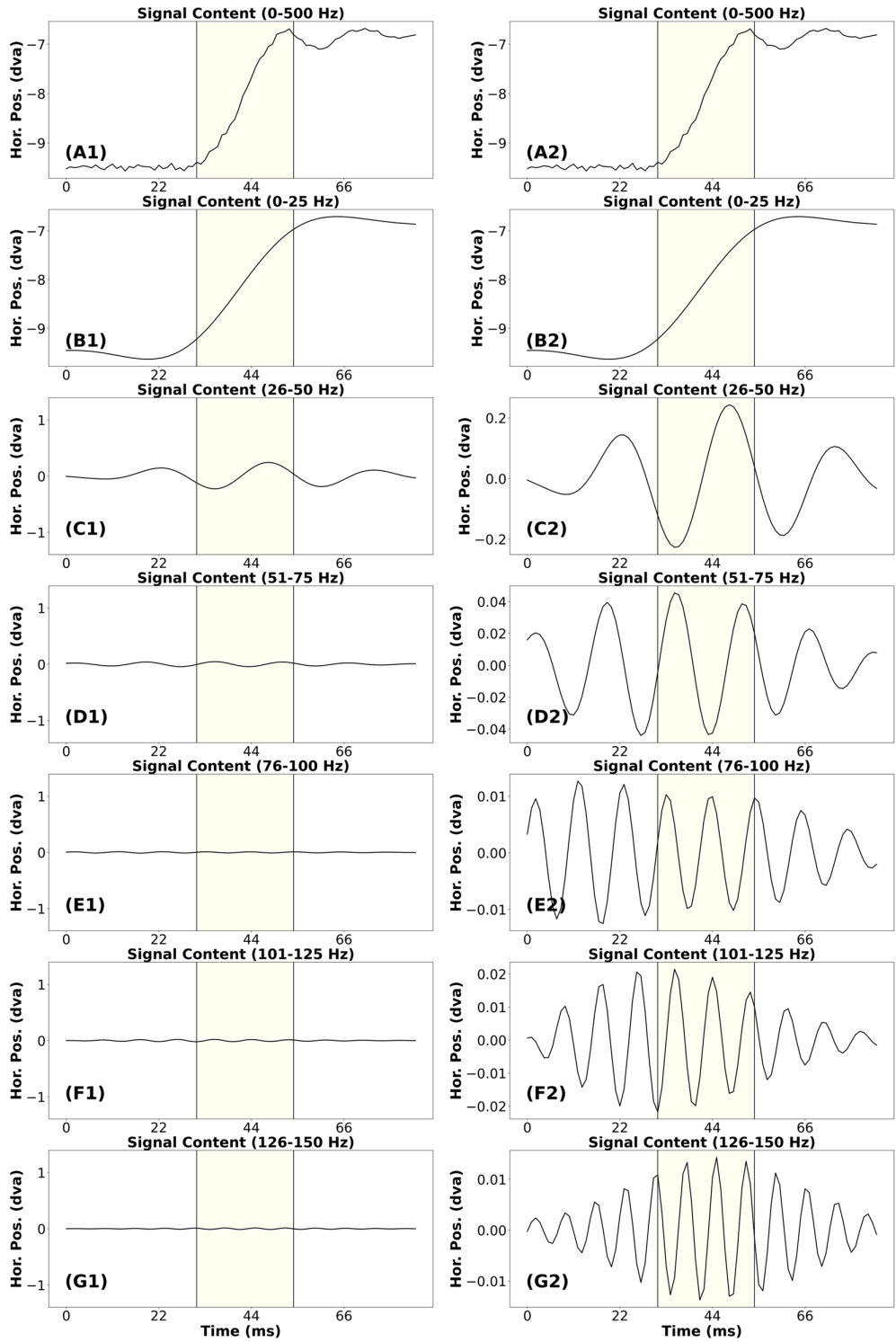

Appendix Figure 1. Signal frequency content analysis of a noisy saccade.

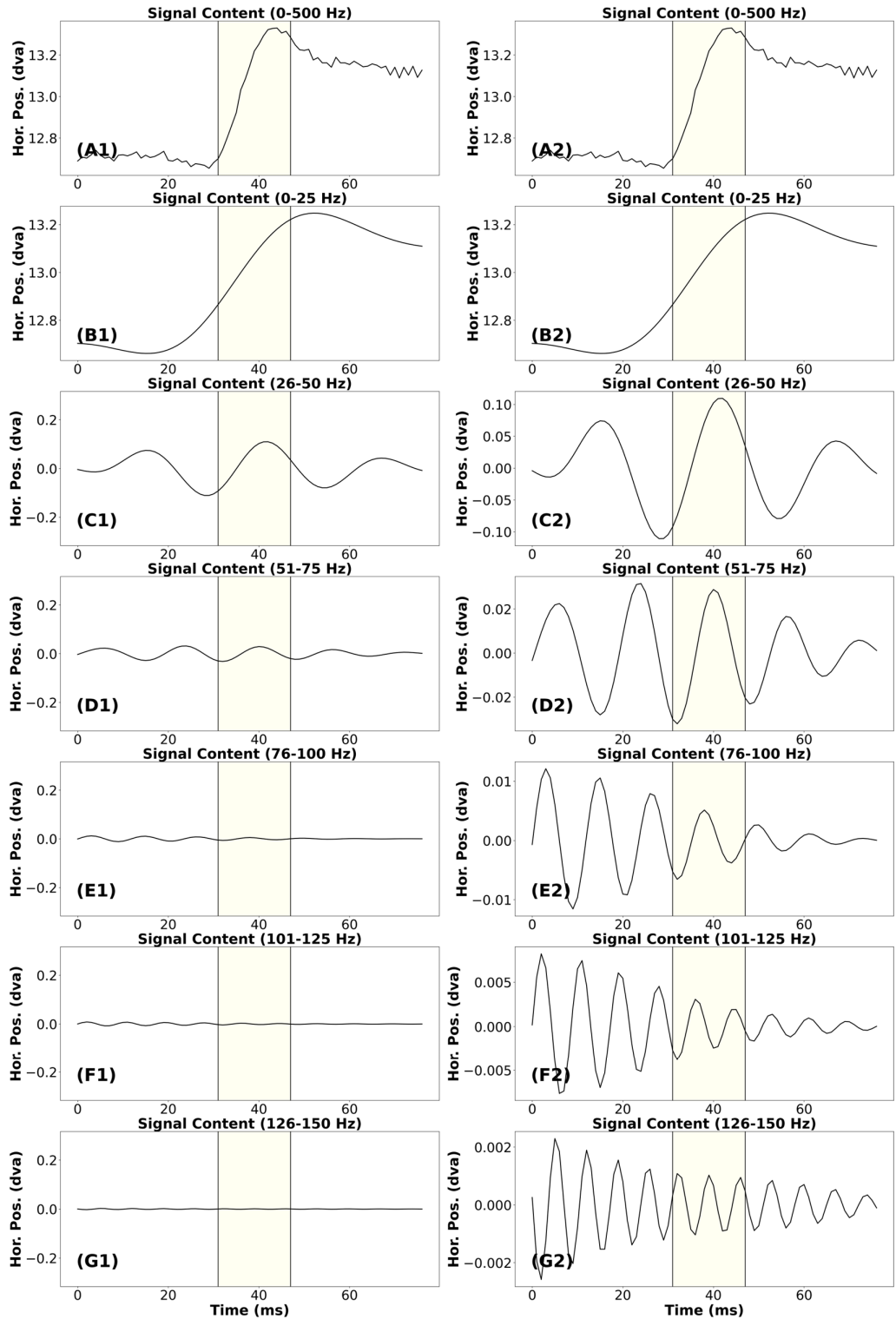

Appendix Figure 2. Signal frequency content analysis of a clean microsaccade.

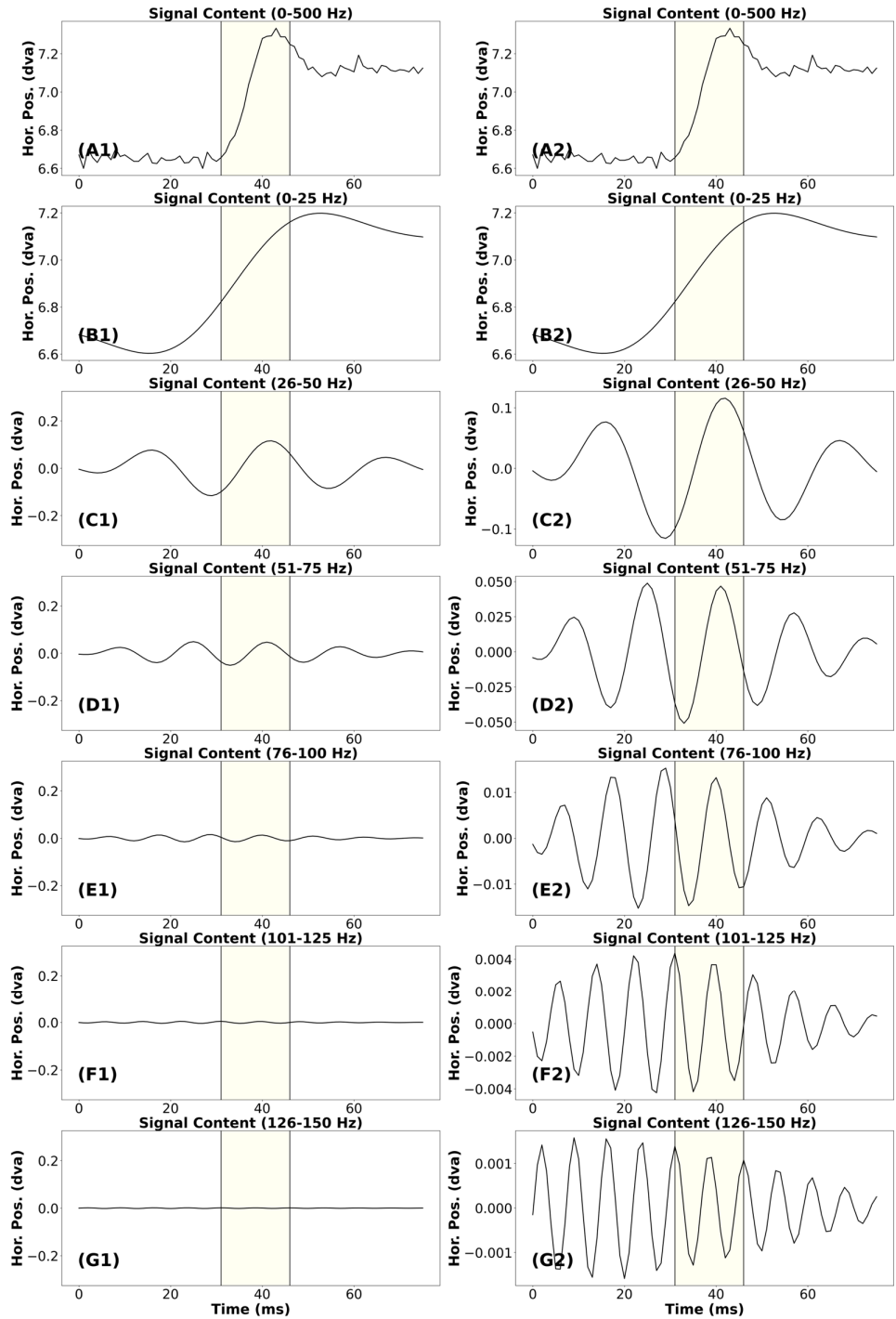

Appendix Figure 3. Signal frequency content analysis of a noisy microsaccade.

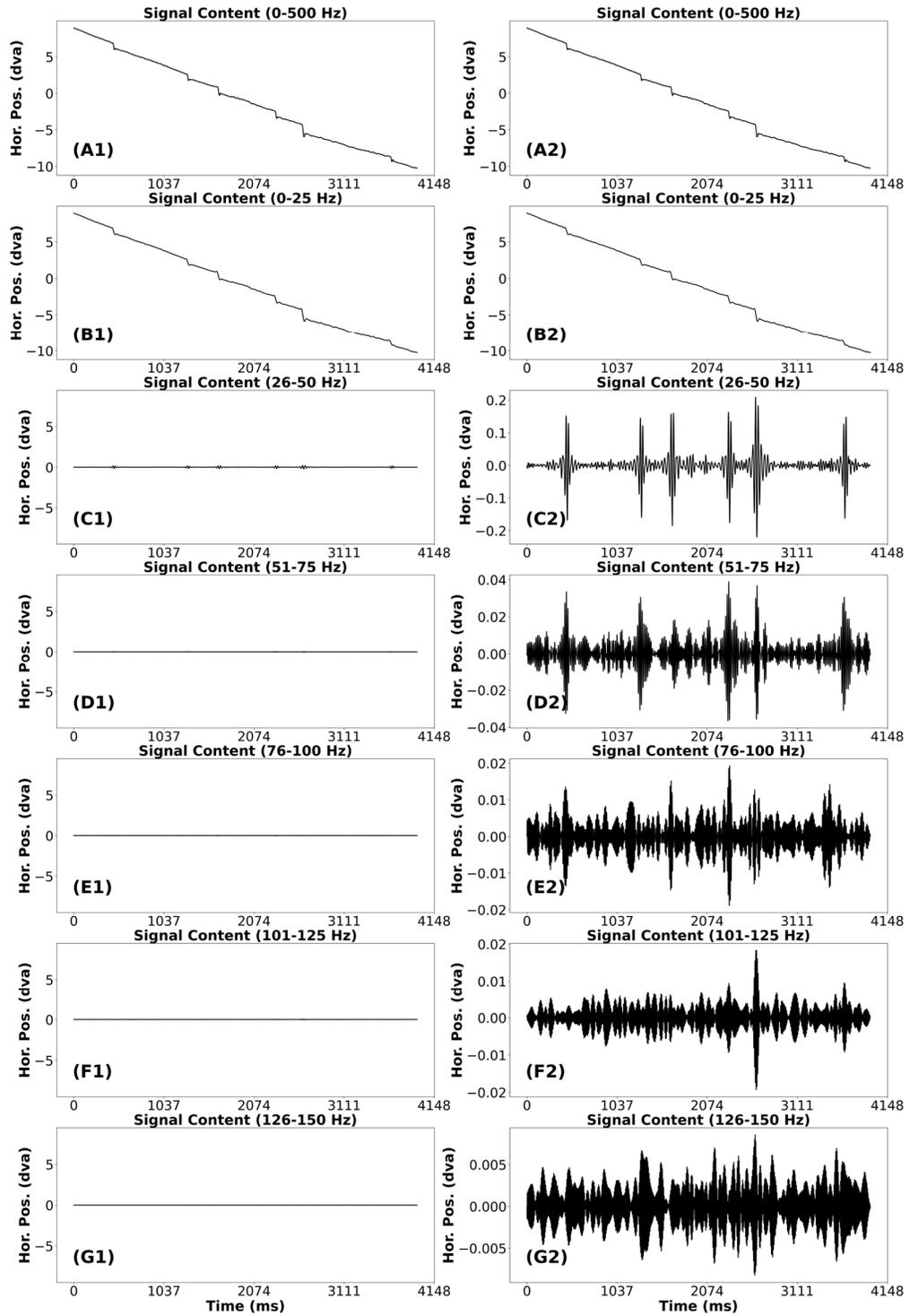

Appendix Figure 4. Signal frequency content analysis of a relatively clean smooth pursuit segment with catch-up saccades.

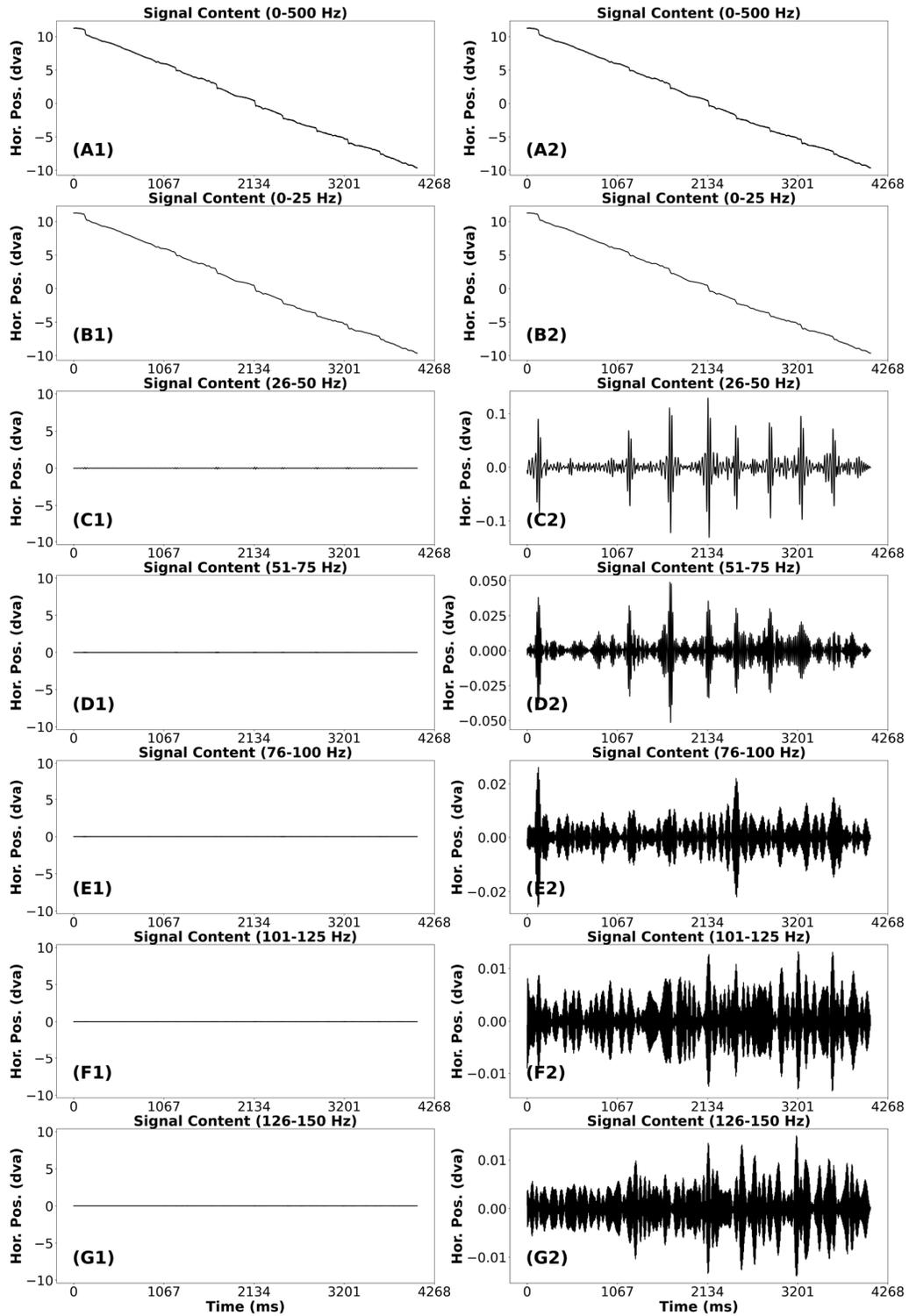

Appendix Figure 5. Signal frequency content analysis of a relatively noisy smooth pursuit segment with catch-up saccades.